\documentclass[%
 reprint,
 amsmath,amssymb,
 aps,nofootinbib,
  twocolumn,
]{revtex4-1}
\usepackage[usenames, dvipsnames]{color}
\usepackage[autostyle]{csquotes}
\usepackage{hyperref}
\usepackage{color}
\usepackage{graphicx}
\usepackage{dcolumn}
\usepackage{bm}
\usepackage{enumerate}
\usepackage{ulem}

\definecolor{MyDarkBlue}{rgb}{0,0.1,0.7}
\hypersetup{colorlinks,breaklinks=true,
  urlcolor={blue},citecolor={MyDarkBlue},linkcolor={Blue}}

\begin{document}


\title{Dynamical aspects of asymmetric Eddington gravity with scalar fields}
\author{Hemza Azri$^{1}$}
\email{hmazri@uaeu.ac.ae; hemza.azri@cern.ch}
\author{Salah Nasri$^{1,2}$}%
\email{snasri@uaeu.ac.ae; salah.nasri@cern.ch}
\affiliation{$^{1}$%
 Department of Physics, United Arab Emirates University,
Al Ain 15551 Abu Dhabi, UAE
}%
\affiliation{$^{2}$%
International Center for Theoretical Physics, Trieste, Italy}%

\date{\today}

\begin{abstract}
In Eddington gravity, the action principle involves only the symmetric parts of the connection and the Ricci tensor, with a metric that emerges proportionally to the latter. Here, we relax this symmetric character, prolong the action with the antisymmetric parts of the Ricci term, and allow for various couplings with scalar fields. We propose two possible invariant actions formed by distinct combinations of the independent Ricci tensors and show that the generated metric must involve an additional antisymmetric part due to the relaxation of the symmetrization property. The comprehensive study shows that the second curvature influences the dynamics of the connection, hence its solution in terms of the metric, and the evolution of the scalar fields. These new dynamical features are expected to stand viable and to have interesting implications in domains where scalar fields are indispensable.      
\end{abstract}

\maketitle

\section{Preliminary remarks and motivation}
General relativity (GR) has revolutionized our understanding of gravity and the structure of the universe at various scales. Despite cosmological puzzles such as the requirement to invoke some dark energy  components, as well as the lack of a complete quantum description of the gravitational interaction, GR remains the only theory that, successfully, relates spacetime geometry to the physical phenomenon. The intimate connection between gravitation and spacetime structure has led to several proposals in attempts to incorporate the other (non-gravitational) interactions into a wider geometric description \cite{unified_field_theory}. This early days unifying endeavor turned out to be unsuccessful, and ended, especially after the enormous progress made in elementary particle physics in which electroweak and strong interactions do not seem to be aspects of the spacetime geometry like gravity. 

Nevertheless, apart from the  motivation behind these early proposals, mainly by Einstein, Schr\"{o}dinger and Eddington \cite{unified_field_theory, eddington, einstein_nature, hlavaty}, one of the interesting ideas that can be drawn from these works is to be less restrictive in choosing the type of the spacetime geometry. In other words, it is  well known that in GR, one assumes \textit{a priori} that the background geometry is fully Riemannian, i.e. spacetime connection becomes torsionless and metric compatible; thus, it is given (also the curvature) in terms of the metric which is the only remaining geometric quantity. That being said, a geometry free of this restriction would involve additional fields such as torsion and non-metricity that would finally bring out interesting effects \cite{lavinia_review}. For instance, one of the most interesting modification of GR that allows spacetime to have torsion is the famed  Einstein-Cartan-Kibble-Sciama theory which relates torsion to the spin density of matter \cite{cartan_kibble1} (see also \cite{cartan_kibble2} for its cosmological implications.) Coupling to higher spin fields in this theory has also been studied \cite{bicak}. 

Now, one raises the question of whether it is the metric (as in GR) or the affine connection which stands as the central object in the variational  principle.

While both connection and metric could be considered fundamental (though independent) in an interesting formulation of gravity named Palatini, there have been, however, attempts to go beyond this formulation where gravity stands only on connections. This idea of purely affine gravity, in which the metric seems to emerge as an integration constant, has been developed by Eddington by restricting the dynamics to only symmetric parts of the connection and the Ricci curvature \cite{eddington}. In Eddington gravity, since the metric concept is not defined, one builds an invariant action via volume measure given by the square-root of the determinant of the Ricci tensor. This was followed by several attempts to extend it by incorporating different type of matter sources \cite{kijowski, treder, demir_matter, cota_matter, azri_separate, azri_immersed, pad}.

In this regard, unlike fermions and gauge bosons which to some extent still require efforts to incorporate them in Eddington gravity completely, scalar fields on the other hand are found to be easily placed in metric-free actions, and allow for interesting features when applied to inflation \cite{affine_inflation, entropy_production_in_affine_inflation}, gravitational dark matter \cite{affine_dark_matter} as well as spontaneous scalarization \cite{scalar_connection_gravity} and other phenomena \cite{inducing_gravity}. These models are still formulated through the strong Eddington's criteria of symmetric connection and Ricci tensor. Therefore, being less restrictive when using these quantities in models of gravity, would be more natural, and one would have interesting and wide setup at hand. 

In this work, we study various dynamical aspects of a purely affine gravity without imposing any restriction on the affine connection and the curvature in particular. We also pay much attention to the dynamics of multiple scalar fields that couple to gravity via the second Ricci tensor. In this study, the metric tensor, which is absent in primary actions, will arise through the dynamical equations, and since the Ricci tensor is not taken symmetric \textit{a priori}, this implies that the metric might involve a non-symmetric part that interacts with matter. Scalar fields are known to serve for various cosmological and particle physics problems, given that the inflationary paradigm, baryogenesis, and dark energy (if dynamical) are already modeled by scalar fields. The framework that we present here will generalize the dynamics of scalar fields nonminimally coupled to gravity, thus leading to new features through their coupling to the novel degree of freedom, namely the skew-symmetric curvature. We carry out a thorough study of the dynamical aspects of this asymmetric affine gravity and point out various possible cosmological implications in domains where scalars are thought to play important roles.      

The rest of this paper is organized  as follows. In the next section we prepare for our setup by providing some preliminary concepts in purely affine spacetime, such as the notion of a general affine connection, curvature, and the construction of the two types Ricci tensors without referring to the concept of metric. In section \ref{section:actions_and_variation}, we involve multiple scalar fields into the study and illustrate how invariant actions are formed based on the relevant second-rank tensors. We then derive the main field equations from a principle of variations in which the field configurations are described by the connections (for the gravitational sector) and the scalars for matter. We also discuss the effects of the curvature on the scalar fields and how to generate the metric tensor. We then conclude in section \ref{section:conclusion}.

\section{Purely affine spacetime: Preliminary concepts}
\label{section:preliminary_concepts}
In the present setup, spacetime is equipped with an asymmetric affine connection $\bm{\Gamma}(x)$ as its central object, from which one defines the covariant derivative $\bm{\nabla}$. 
In general, the primary role of an affine connection is to map the tangent space at a point $P$ to that of a neighboring point $P + dP$, which  is performed by parallel transfering a vector $\bm{\xi}^{\alpha}(x)$ via the coordinate displacement $\delta x^{\nu}$ as
\begin{eqnarray}
\delta \bm{\xi}^{\alpha} = -\bm{\Gamma}^{\alpha}_{\,\,\mu\nu}\bm{\xi}^{\mu}\delta x^{\nu}.
\end{eqnarray}

A geometry with this simple structure is said to be purely affine, and the concept of metric in GR is not defined yet. Therefore, the so-called geodesics, i.e. the straightest lines in this geometry, are not said to extremize distances between points, but are defined as those curves with tangent vectors that remain parallel to themselves through parallel displacements.

The curvature tensor in this case  which generalizes the Riemann tensor of GR, arises as a measure of the failure of the commutativity of the covariant derivation of any vector field $\bm{\xi}^{\alpha}(x)$ as \cite{lovelock, azri_thesis} 
\begin{eqnarray}
\left[\bm{\nabla}_{\mu}, \bm{\nabla}_{\nu} \right]\bm{\xi}^{\alpha}
= \bm{R}^{\alpha}_{\,\, \lambda \mu \nu}\,\bm{\xi}^{\lambda}
-2 \bm{S}^{\lambda}_{\,\,\mu\nu} \bm{\nabla}_{\lambda}\bm{\xi}^{\alpha},
\end{eqnarray}
where $\bm{S}^{\lambda}_{\,\,\mu\nu}$ is the torsion tensor (the antisymmeric part of the affine connection) and the spacetime curvature tensor is given in terms of the connection as 
\begin{eqnarray}
\label{riemann}
\bm{R}^{\alpha}_{\,\, \lambda \mu \nu}=
\partial_{\mu}\bm{\Gamma}^{\alpha}_{\,\,\lambda \nu}
-\partial_{\nu}\bm{\Gamma}^{\alpha}_{\,\,\lambda \mu}
+\bm{\Gamma}^{\alpha}_{\,\,\rho \mu} \bm{\Gamma}^{\rho}_{\,\,\lambda \nu}
-\bm{\Gamma}^{\alpha}_{\,\,\rho \nu} \bm{\Gamma}^{\rho}_{\,\,\lambda \mu}
\end{eqnarray}
This curvature allows for the construction of two Ricci tensors which are obtained by contraction
\begin{eqnarray}
\label{r_and_q}
R_{\mu \nu}(\Gamma)
\equiv
\bm{R}^{\lambda}_{\,\, \mu \lambda \nu}, \quad \text{and} \quad
Q_{\mu\nu}(\Gamma) \equiv
\bm{R}^{\lambda}_{\,\, \lambda \mu \nu}.
\end{eqnarray}
In terms of the affine connection, the second Ricci tensor takes a simple form
\begin{eqnarray}
\label{second_type_ricci}
Q_{\mu\nu}(\Gamma)=
\partial_{\mu} \bm{\Gamma}^{\lambda}_{\,\,\lambda \nu}
-\partial_{\nu} \bm{\Gamma}^{\lambda}_{\,\,\lambda \mu}.
\end{eqnarray}
For  arbitrary connection, these two Ricci curvatures are totally independent. In fact, from the symmetry properties of the curvature tensor (\ref{riemann}), one can show that \cite{lovelock}
\begin{eqnarray}
Q_{\mu\nu}(\Gamma) =
&& R_{\nu\mu}(\Gamma) -R_{\mu\nu}(\Gamma) \nonumber \\
&&+\bm{\nabla}_{\alpha}\bm{S}^{\alpha}_{\,\,\mu\nu}
+\bm{\nabla}_{\nu}\bm{S}_{\mu}
-\bm{\nabla}_{\mu}\bm{S}_{\nu} 
+\bm{S}_{\alpha}\bm{S}^{\alpha}_{\,\,\mu\nu},
\end{eqnarray}
where $\bm{S}_{\mu}=\bm{S}^{\alpha}_{\,\,\mu\alpha}$ is the torsion vector. However, this identity shows that if the affine connection is symmetric (when torsion vanishes), then the second Ricci curvature (\ref{second_type_ricci}) is nothing but the skew-symmetric part of the Ricci curvature $R_{\mu\nu}(\Gamma)$. This means that, generally, the antisymmetric part of the affine curvature must play an important role in the dynamics unless it is assumed to be zero from scratch as in the case of Eddington gravity \cite{eddington}. In the subsequent sections, we will propose invariant actions that involve both Ricci tensors coupled to multiple scalar fields.

\section{Variational principle in asymmetric Eddington gravity}
\label{section:actions_and_variation}
In addition to the geometric sector which has been set up in the previous section, we consider  matter  represented by scalar fields  $\bm{\phi}^{a}$ ($a =1,\dots, N$). The main aim of this section is to describe and study the possible coupling of these scalar fields with the spacetime curvatures (\ref{r_and_q}), thus, in a purely affine framework. Variational principle of field theory in curved spacetime is known to be constructed from Lagrangian (scalar) densities formed by contractions of relevant tensors. This process requires a metric at the first place to allow for lowering and raising indices. This is not possible at this stage since the geometry discussed above is metric-free. One possibility that goes beyond the familiar field theoretic polynomial Lagrangian is to use the square-root of the determinant  of two-rank tensors. Therefore, the following important steps are required in constructing an invariant action in the previous setup
\begin{enumerate}[(i)]
    \item For the gravitational sector (geometric part), the natural  quantities  to mind are the Ricci tensors (\ref{r_and_q}). Hence, the first quantity to consider is  a linear combination of these tensors.
   \item On the other hand, scalar fields in particular  do not require a metric so that one gets a two-rank tensors $\partial_{\mu}\bm{\phi}^{a}\partial_{\nu}\bm{\phi}^{a}$ generating their kinetric terms. 
   \item Scalar potential energies can enter the invariant action rendering the total quantity dimensionless. The latter requires that the potential must come in division instead of addition. 
   
   Hence, unlike the familiar actions, the total quantity may become singular if the total potential vanishes, a case which occurs in spontaneous symmetry breaking models when the fields stay at their minimum. Nevertheless, nonzero potentials are recovered in this type of potentials by adding a cosmological constant term that guarantees the existence of a nonzero vacuum energy \cite{affine_inflation}. 
   \item The final point concerns whether the scalar fields couple directly to the spacetime curvature through only  one of  the Ricci tensors in (\ref{r_and_q}) or both of them. In either cases, we say that matter is coupled nonminimally to (purely affine) gravity.
\end{enumerate}

These steps and requirements have been considered in forming models of affine inflation in the symmetric case, i.e. when the spacetime connection is torsionless and only the symmetric part of the Ricci curvature is used \cite{affine_inflation, entropy_production_in_affine_inflation}. 

In what follows, based on the above points, we generalize these models by keeping the spacetime geometry free of these restrictions. 

\subsection{The case of \texorpdfstring{$R_{\mu\nu}(\Gamma) + \bm{\lambda}\, Q_{\mu\nu}(\Gamma)$}{R_{\mu\nu}(\Gamma) + \lambda Q_{\mu\nu}(\Gamma)}}


First of all, it is worth mentioning that as in the case of GR or standard field theory in curved space, there is no unique Lagrangian density. However, the first model of interest that satisfies the above properties is merely based on a linear combination of the Ricci tensors (\ref{r_and_q}). Now, the scalar fields will manifest through their derivatives which in turn can be involved to extend the two tensor formed by the previous linear combination. An important coupling (see point (iv) above) comes from the  interaction of the scalar fields with this linear combination. 
All this can be expressed   by the invariant action  
\begin{eqnarray}
\label{general_action}
S[\Gamma, \bm{\phi}]=
\int d^{4}x \frac{\sqrt{ |f(\bm{\phi}) \left(R_{\mu\nu} + \bm{\lambda}\, Q_{\mu\nu} \right)  - \partial_{\mu}\bm{\phi}^{a} \partial_{\nu}\bm{\phi}^{a}|}}
{V(\bm{\phi})} \nonumber \\
\end{eqnarray}
where $\bm{\lambda}$ is a dimensionless constant, $f(\bm{\phi})$ is a nonminimal coupling function,  and $V(\bm{\phi})$ is a nonzero potential.

For brevity, we  denote the total rank-two tensor, from which this action  is formed, as 
\begin{eqnarray}
\label{tensor_k1}
K_{\mu\nu}(\Gamma,\bm{\phi})\equiv
f(\bm{\phi}) \left(R_{\mu\nu} + \bm{\lambda}\, Q_{\mu\nu} \right)  - \partial_{\mu}\bm{\phi}^{a} \partial_{\nu}\bm{\phi}^{a}.
\end{eqnarray}
Since the affine connection is arbitrary, variations of the first Ricci curvature will merely involve the torsion tensor as
\begin{eqnarray}
\delta R_{\mu\nu}=
\bm{\nabla}_{\alpha}\left(\delta \bm{\Gamma}_{\,\,\mu\nu}^{\alpha} \right)
- \bm{\nabla}_{\nu}\left(\delta \bm{\Gamma}_{\,\,\mu\alpha}^{\alpha} \right)
-2\bm{S}^{\alpha}_{\,\,\beta \nu} \delta \bm{\Gamma}^{\beta}_{\,\,\mu\alpha}.
\end{eqnarray}

Before performing the variation of the action, it is necessary to bring here some useful formulae to manipulate any surface term that will appear through the variation. The covariant derivation of any scalar density $\mathfrak{E}$ (of weight +1), like the one inside integral (\ref{general_action}), reads
\begin{eqnarray}
\label{gauss}
\bm{\nabla}_{\mu}\mathfrak{E} =
\partial_{\mu}\mathfrak{E} - \bm{\Gamma}^{\alpha}_{\,\,\alpha \mu} \mathfrak{E}.
\end{eqnarray}
With the aid of this identity, one states the Gauss theorem where for a given vector (or tensor) field $\bm{\xi}^{\mu}$ we have 
\begin{eqnarray}
\label{surface_term}
\int d^{4}x \bm{\nabla}_{\mu}\left(\mathfrak{E}\, \bm{\xi}^{\mu} \right)=
2\int d^{4}x \left(\mathfrak{E}\, \bm{\xi}^{\mu}\right)\bm{S}_{\mu}.
\end{eqnarray}
Here, one notices the familiar surface term (vanishing of the integral) when the connection is symmetric. 

The field equations of interest are now obtained by extremizing  the action (\ref{general_action}) with respect to the arbitrary connection $\bm{\Gamma}$,  i.e. $\delta_{_{\bm{\Gamma}}}S =0$, which  yields
\begin{widetext}
\begin{eqnarray}
\label{general_dynamical_equation}
&&
\bm{\nabla}_{\alpha} \left(f(\bm{\phi})\frac{\sqrt{|K|}}{V(\bm{\phi})} (K^{-1})^{\mu\nu} \right) 
- \bm{\nabla}_{\beta} \left(f(\bm{\phi})\frac{\sqrt{|K|}}{V(\bm{\phi})} (K^{-1})^{\mu\beta} \right)\delta^{\nu}_{\,\,\alpha} 
+2f(\bm{\phi})\frac{\sqrt{|K|}}{V(\bm{\phi})} (K^{-1})^{\mu\beta}\bm{S}_{\,\,\alpha\beta}^{\nu} 
\nonumber \\
&&-2f(\bm{\phi})\frac{\sqrt{|K|}}{V(\bm{\phi})} (K^{-1})^{\mu\nu}\bm{S}_{\alpha} 
+2f(\bm{\phi})\frac{\sqrt{|K|}}{V(\bm{\phi})} (K^{-1})^{\mu\beta}\bm{S}_{\beta} \delta^{\nu}_{\,\,\alpha} 
-2\bm{\lambda}\, \partial_{\beta} \left(f(\bm{\phi})\frac{\sqrt{|K|}}{V(\bm{\phi})} (K^{-1})^{[\nu\beta]} \right)\delta^{\mu}_{\,\,\alpha} =0.
\end{eqnarray}
\end{widetext}

These equations represent the gravitational field equations in the purely affine formulation. At first sight, they seem quite complicated and impossible to solve directly since they involve high order terms in the connection, particularly through the determinant of the tensor field $K(\bm{\Gamma},\bm{\phi})$. The way to deal with this issue is to follow Eddington's approach and write these equations in terms of a second-rank tensor that will appear as an integration constant. This new quantity would simply lead to the so-called metric tensor from which the equations of motion take a familiar form. We now turn to the transition from purely affine dynamics to the metric structure.  

\subsubsection{Metric structure and field equations}
As we have seen so far, unlike GR, the metric tensor in purely affine gravity is not assumed from the beginning. In Eddington gravity, since the action is constructed out of the symmetric part of the Ricci curvature, the metric (necessarily symmetric) emerges easily as a quantity proportional to the Ricci tensor, and the theory leads to Einstein's gravity with a nonzero cosmological constant. Now, since the assumption of this symmetric character is relaxed, a general tensor must be defined so that it extends the metric of Eddington gravity.  To that end, we introduce an invertible two-rank tensor $\bm{g}_{\mu\nu}(x)$ and define an associated tensor density as
\begin{eqnarray}
\label{metric_density}
M^{2}\sqrt{|\bm{g}|}\,\bm{g}^{\mu\nu} =
f(\bm{\phi})\frac{\sqrt{|K(\bm{\Gamma},\bm{\phi})|}}{V(\bm{\phi})} (K^{-1})^{\mu\nu},
\end{eqnarray}
where we have introduced a constant mass scale $M$ for dimensional homogeneity of the equation. The form of this tensor density is chosen so that the dynamics coincides with the symmetric models, i.e, the limiting case where $\bm{g}_{\mu\nu}$ is reduced to the metric tensor and the resulting field equations describe the gravitational equations with a nonminimally coupled scalar fields \cite{affine_inflation, entropy_production_in_affine_inflation}. Additionally, identity (\ref{metric_density}) can be seen as a generalized metrical density of Eddington gravity in free space which is defined as the derivative of the Lagrangian with respect to the symmetric part of the Ricci tensor \cite{eddington}  
\begin{eqnarray}
\label{metric_as_derivative_wrt_ricci}
\sqrt{|\bm{g}|}\,\bm{g}^{\mu\nu} \equiv
2\frac{\partial \mathcal{L}}{\partial R_{\mu\nu}},
\end{eqnarray}
and appears like a momentum conjugate of the field configuration (considered as the affine connection \cite{kijowski}). Therefore, when the Ricci tensor in (\ref{metric_as_derivative_wrt_ricci}) is symmetric, the generated metric is automatically symmetric and will  describe the physical metric (gravitational field).

Returning to our case, now with the new quantity (\ref{metric_density}), the dynamical equation (\ref{general_dynamical_equation}) is reduced to a linear equation for the connection $\bm{\Gamma}$
\begin{eqnarray}
\label{metric_dynamical_equation}
&&\bm{\nabla}_{\alpha}(\sqrt{|\bm{g}|}\,\bm{g}^{\mu\nu})
+2\sqrt{|\bm{g}|}\,\bm{g}^{\mu\beta}\left(\bm{S}^{\nu}_{\,\,\alpha\beta} + \frac{1}{3}\bm{S}_{\beta} \delta^{\nu}_{\,\,\alpha} \right) \nonumber \\
&&-2\sqrt{|\bm{g}|}\,\bm{g}^{\mu\nu}\bm{S}_{\alpha} 
+2\bm{\lambda}\left(\frac{1}{3}\mathcal{J}^{\mu} \delta^{\nu}_{\,\,\alpha}- \mathcal{J}^{\nu} \delta^{\mu}_{\,\,\alpha}
\right) =0,
\end{eqnarray}
where we have defined the vector density
\begin{eqnarray} 
\mathcal{J}^{\mu}= \partial_{\alpha}\left(\sqrt{|\bm{g}|}\bm{g}^{[\mu \alpha]}\right).
\end{eqnarray} 

From (\ref{metric_density}), it is clear that the tensor $\bm{g}_{\mu\nu}$ is not symmetric, and  thus, its skew-symmetric part must also have crucial effects on the dynamics. A useful equation that constraints the antisymmetric part of this tensor can be obtained from (\ref{general_dynamical_equation}) by the contraction $\alpha=\mu$, which yields 
\begin{eqnarray}
\label{divergence_of_g}
(1 + 4\bm{\lambda})\mathcal{J}^{\mu}=0.
\end{eqnarray}
Hence, its divergence can be determined from the theory unless $\bm{\lambda}= -1/4$. The latter which appears also in the vacuum case \cite{murphy, poplawski}, is a result of the invariance of action (\ref{general_action}) under projective transformation, $\bm{\Gamma}^{\alpha}_{\,\,\mu\nu} \rightarrow \bm{\Gamma}^{\alpha}_{\,\,\mu\nu} + \delta^{\alpha}_{\,\, \mu}\bm{\xi}_{\nu}$, for an arbitrary real one-form field $\bm{\xi}_{\mu}(x)$. In fact, the linear combination of the two Ricci tensors transforms as
\begin{eqnarray}
R_{\mu\nu} +\bm{\lambda}\, Q_{\mu\nu} \rightarrow
R_{\mu\nu} +\bm{\lambda}\, Q_{\mu\nu}
+2(1 +4\bm{\lambda})(\partial_{\mu}\bm{\xi}_{\nu}-\partial_{\nu}\bm{\xi}_{\mu}) \nonumber \\
\end{eqnarray}

In addition to general coordinate transformations, several gravity models which are based on the affine Ricci curvature enjoy also the projective transformation. The well-known feature of this transformation is that it does not affect the rule of parallel displacement of vectors along arbitrary curves.

Therefore, yet the case $\bm{\lambda}= -1/4$ which spoils the constraint on $\mathcal{J}^{\mu}$, is preserved. It is not altered by the presence of the scalar fields given that the tensor density (\ref{metric_density}) is a relevant generalization of the metric in the vacuum case.

Notice also that, for arbitrary values of $\bm{\lambda}$, both Ricci terms are already invariant under the projective symmetry when the one-form $\bm{\xi}_{\mu}(x)$ is a gradient. Needless to say, it was usually realized that the dynamics of the second Ricci tensor resembles to some extent that of a complex vector field. The constraint (\ref{divergence_of_g}) reminds us of a massive vector field where a divergence-free relation arises only when the mass of the field is nonzero, and in the massless case, one ends up with an additional symmetry; the gauge invariance.

Returning to identity (\ref{metric_density}) which can be easily inverted to finally take the form  
\begin{eqnarray}
\label{field_equations_1}
f(\bm{\phi}) \left(R_{\mu\nu} + \bm{\lambda}\, Q_{\mu\nu} \right)  - \partial_{\mu}\bm{\phi}^{a} \partial_{\nu}\bm{\phi}^{a} 
-\frac{M^{2}V(\bm{\phi})}{f(\bm{\phi})} \bm{g}_{\mu\nu} =0 \nonumber \\
\end{eqnarray}
This equation has the form of a gravitational field equations where the spacetime connection that defines the curvature satisfies the non trivial dynamical equation (\ref{metric_dynamical_equation}). Therefore, solving the latter in terms of the tensor field $\bm{g}$ and the torsion, will provide the complete gravitational dynamics. One may also write these equations in a standard form by constructing a generalized Einstein tensor and obtains
\begin{eqnarray}
\label{einstein_field_equations_1}
&&
f(\bm{\phi})\left(R_{\mu\nu} - \frac{1}{2}\bm{g}_{\mu\nu}\bm{g}^{\alpha\beta} R_{\alpha\beta}\right) \nonumber \\
&&=
\partial_{\mu}\bm{\phi}^{a} \partial_{\nu}\bm{\phi}^{a}
-\frac{1}{2}\bm{g}_{\mu\nu}
\bm{g}^{\alpha\beta}
\partial_{\alpha}\bm{\phi}^{a} \partial_{\beta}\bm{\phi}^{a}
-\frac{M^{2}V(\bm{\phi})}{f(\bm{\phi})} \bm{g}_{\mu\nu} \nonumber \\
&&-\bm{\lambda}f(\bm{\phi})\left(Q_{\mu\nu}
-\frac{1}{2}\bm{g}_{\mu\nu}\bm{g}^{\alpha\beta} Q_{\alpha\beta}\right).
\end{eqnarray}
A few important aspects of these equations are worth stating here:
\begin{itemize}
    \item The scalar fields interact with both the asymmetric metric and the two types of the curvature,  and hence, the equations generalize the case of nonminimal coupling to gravity. In this case, notice the absence of the second derivatives of the scalar fields compared to familiar models with nonminimal couplings, and the reason here is simply that action (\ref{general_action}) is linear in the connection. As a result, one can easily show that the nonminimal coupling terms $f(\bm{\phi})$ could be completely absorbed by redefining the potential and the kinetric terms of the scalars without altering the geometric sector by any transformation like the familiar conformal mapping. Inflationary models driven by multi-scalar fields are known to predict a second type of cosmological perturbations, namely isocurvature \cite{multifield_inflation}, especially when the fields are nonminimally coupled to gravity \cite{entropy_production_in_affine_inflation, kaiser}. The present setup offers possible effects on these generic predictions due to the interaction of the multifields with the new degrees of freedom (curvature) through the last term of the gravitational equations (\ref{einstein_field_equations_1}).
    \item In addition to the second curvature term, the other important element which is also present due to the relaxation of the symmetric character is the torsion tensor field. Unlike the curvature, this tensor does not appear explicitly in the equations of motion (\ref{einstein_field_equations_1}) but should emerge through the connection that defines the curvature. Its effect on the dynamics is to shift the curvature and interacts with the scalar fields. Here, a particular limiting case which is in the spirit of Eddington gravity arises when only the connection (not the curvature) is symmetric. As stated in section \ref{section:preliminary_concepts}, the two Ricci tensors are not independent in this case since we simply would have $Q_{\mu\nu}(\Gamma)=2R_{[\mu\nu]}(\Gamma)$.
    \item The final remark  is that this setup cannot be considered as only a different formulation to gravity. Indeed, the symmetric cases, such as Eddington gravity and its extensions with scalar fields are admitted as purely affine formulations to GR despite their new cosmological features \cite{kijowski, affine_inflation, entropy_production_in_affine_inflation}. While these models can have their GR (purely metric) counterparts, the present work however necessitates the concept of an affine connection in the first place for the other degrees of freedom (second Ricci tensor and torsion) to make sense.
\end{itemize}
 
The gravitational field equations need to  be accompanied with an equation for the evolution  of the scalar fields, which can be  derived by performing a variation of the action (\ref{general_action}) with respect to $\bm{\phi}^{a}$, and obtain
\begin{eqnarray}
&&\partial_{\nu} \left(\frac{\sqrt{|K|}}{V(\bm{\phi})} K^{(\mu\nu)}\partial_{\mu} \bm{\phi}^{a} \right)
+\frac{f_{,a}}{2}\frac{\sqrt{|K|}}{V(\bm{\phi})} K^{\mu\nu}\left(R_{\mu\nu}+\bm{\lambda} Q_{\mu\nu}\right)
\nonumber \\
&&-\frac{\sqrt{|K|}}{V^{2}(\bm{\phi})}V_{,a}=0,
\end{eqnarray}
where the underscript "," represents the derivative with respect to $\bm{\phi}^{a}$. We notice the absence of the torsion compared to the field equations (\ref{general_dynamical_equation}) since the derivative operators acting on the scalars are only ordinary, and thus when integrating by parts the surface terms would not involve the torsion vector as in (\ref{surface_term}). Nevertheless, as we have mentioned above, the torsion is implicitly present through the asymmetric connection in the curvature terms. 

With the defined tensor density (\ref{metric_density}), this equation finally reads
\begin{eqnarray}
\label{box_phi_1}
&&\frac{\partial_{\nu}\left(\sqrt{|\bm{g}|}\,\bm{g}^{(\mu\nu)} \partial_{\mu}\bm{\phi}^{a} \right)}{\sqrt{|\bm{g}|}}
-\frac{\partial V}{\partial \bm{\phi}^{a}}
+\frac{1}{2}\frac{\partial f}{\partial \bm{\phi}^{a}}
\bm{g}^{\mu\nu}R_{\mu\nu} \nonumber \\
&&
+\frac{\bm{\lambda}}{2}\frac{\partial f}{\partial \bm{\phi}^{a}}
\bm{g}^{\mu\nu}Q_{\mu\nu}
+ \bm{\Psi}(\bm{\phi}^{a}) =0,
\end{eqnarray}
where the last quantity is give by
\begin{eqnarray}
\label{psi_1}
\bm{\Psi}(\bm{\phi}^{a})=
\left(1-\frac{M^{2}}{f} \right)\frac{\partial V}{\partial \bm{\phi}^{a}}
-\frac{1}{f}\frac{\partial f}{\partial \bm{\phi}^{a}}\bm{g}^{(\mu\nu)}
\partial_{\mu}\bm{\phi}^{a} \partial_{\nu}\bm{\phi}^{a}
\nonumber \\
\end{eqnarray}

Like the gravitational field equations (\ref{einstein_field_equations_1}), this equation has also the standard and familiar (metrical) form. The important quantities in this equation are as follows: (\textit{i}) The first term is nothing but the generalized d'Alembert wave propagator $\Box_{\bm{g}} \bm{\phi}^{a}$ (\textit{ii}) The third term caused by the direct interaction of the scalars with the curvature and it is generic in all theories with nonminimal couplings, though in this case it is the spacetime connection (not the metric) which defines the type of this coupling. An unfamiliar quantity arises in the fourth term which shows the novel nonminimal coupling between matter and curvature (second Ricci tensor) (\textit{iii}) The last term given explicitly in (\ref{psi_1}) is not a result of the relaxation of the symmetric character of the connection and the curvature but it appears even in the symmetric models with nonminimal couplings \cite{affine_inflation, entropy_production_in_affine_inflation}. This quantity would disappear if the fields were coupled only minimally, i.e. when $f \rightarrow M^{2}$. In this case the mass scale is reduced to the Planck mass.

\subsubsection{Dynamics of the connection}
The above dynamics depends on the arbitrary connection which must be determined in terms of the tensor $\bm{g}_{\mu \nu}$ from which one can define a symmetric physical metric. Before going further, one can first simplify the dynamical equation (\ref{metric_dynamical_equation}), based on the projective transformation discussed above. For that, one introduces the following connection 
\begin{eqnarray}
\tilde{\bm{\Gamma}}_{\,\,\mu\nu}^{\alpha}= \bm{\Gamma}_{\,\,\mu\nu}^{\alpha}
+\frac{2}{3}\delta^{\alpha}_{\,\,\mu}\bm{\Gamma}_{\,\,\nu\rho}^{\rho},
\end{eqnarray}
from which arises $\tilde{\bm{\Gamma}}_{\,\,[\mu\rho]}^{\rho}=0$, leading  to considerable simplification as we shall see below.

In terms of this connection, equation (\ref{metric_dynamical_equation}) reads 
\begin{eqnarray}
\label{start_affinity_equation}
\partial_{\alpha}&&\left(\sqrt{|\bm{g}|}\bm{g}^{\mu\nu} \right)
+ \sqrt{|\bm{g}|}\bm{g}^{\rho\nu}\tilde{\bm{\Gamma}}_{\,\,\rho\alpha}^{\mu}
+ \sqrt{|\bm{g}|}\bm{g}^{\mu\beta}\tilde{\bm{\Gamma}}_{\,\,\alpha\beta}^{\nu} \nonumber \\
&&- \sqrt{|\bm{g}|}\bm{g}^{\mu\nu}\tilde{\bm{\Gamma}}_{\,\,\alpha\beta}^{\beta} =2\bm{\lambda}\left(\frac{1}{3}\mathcal{J}^{\mu} \delta^{\nu}_{\,\,\alpha}- \mathcal{J}^{\nu} \delta^{\mu}_{\,\,\alpha}
\right).
\end{eqnarray}

Therefore, the dynamics of model (\ref{general_action}) is summarized in this equation for the connection. A solution of this equation in terms of the tensor  $\bm{g}_{\mu \nu}$ would recast the gravitational field equation (\ref{einstein_field_equations_1}) and the scalar field equations (\ref{box_phi_1}) to finally take a metrical form. 

Relaxation of the symmetric character of the connection and Ricci tensor increases the degrees of freedom of the system to 64 unknown coefficients of $\bm{\Gamma}$, which must be determined in terms of the 16 components of $\bm{g}_{\mu \nu}$. To solve the above equation, we introduce the symmetric and the anti-symmetric parts of the tensor $\bm{g}_{\mu \nu}$ which will be denoted, respectively, as
\begin{eqnarray}
\bm{g}_{(\mu\nu)}\equiv g_{\mu\nu}(x) \quad \text{and} \quad
\bm{g}_{[\mu\nu]}\equiv \mathfrak{f}_{\mu\nu}(x)
\end{eqnarray}
Thus, the case $\bm{\lambda} \neq -1/4$ implies $\mathcal{J}^{\mu}=0$ (see constraint (\ref{divergence_of_g})), and then one can show that the general solution of the dynamical equation (\ref{start_affinity_equation}) is given as \cite{murphy, hlavaty}  
\begin{eqnarray}
\label{solution_for_the_connection}
\tilde{\bm{\Gamma}}_{\,\,\mu\nu}^{\alpha}=&& \Gamma_{\,\,\mu\nu}^{\alpha}(g) + 
\frac{1}{2}\mathfrak{T} ^{\alpha}_{\,\,\mu\nu} +
\mathfrak{f}^{\beta}_{\,\,[\nu}\mathfrak{T}^{\sigma}_{\,\,\mu]\beta}\mathfrak{f}^{\alpha}_{\,\,\sigma} \nonumber \\
&&+g^{\sigma\alpha}\mathfrak{f}_{\rho (\kappa}
\mathfrak{T}^{\rho}_{\,\,\tau)\gamma}
\Big\{ \delta^{\gamma \tau \kappa}_{\sigma (\mu\nu)}
- 2 \delta^{\tau}_{\sigma}\mathfrak{f}^{\gamma}_{\,\,(\mu} \mathfrak{f}^{\kappa}_{\,\,\nu)}
- 2 \delta^{\kappa}_{(\mu}\mathfrak{f}^{\tau}_{\,\,\nu)} \mathfrak{f}^{\gamma}_{\,\,\sigma}
\Big\} \nonumber \\
\end{eqnarray}
where $\Gamma_{\,\,\mu\nu}^{\alpha}(g)$ is the Levi-Civita connection of the metric tensor $g_{\mu\nu}$, and the tensor $\mathfrak{T}$ is given in terms of the covariant derivative with respect to $\Gamma_{\,\,\mu\nu}^{\alpha}(g)$ (denoted $\nabla$) as
\begin{eqnarray}
\mathfrak{T}_{\alpha \mu \nu}=
\nabla_{\alpha}\mathfrak{f}_{\nu\mu}
+\nabla_{\mu}\mathfrak{f}_{\alpha\nu}
+\nabla_{\nu}\mathfrak{f}_{\alpha\mu}.
\end{eqnarray}

Here, raising and lowering the indices are performed by the symmetric metric $g_{\mu \nu}$ which is considered now as a relevant tensor for the gravitational field as in GR. To that end, the spacetime connection is given in terms of the Levi-Civita connection (metric compatible) of $g(x)$. This nontrivial solution would lead to significant deviations from GR when substituted into the field equations (\ref{einstein_field_equations_1}) and (\ref{box_phi_1}).

\subsection{General nonminimal coupling dynamics}
Given the various ways in defining an invariant action, it is clear that action (\ref{general_action}) is not the unique extension of Eddington gravity with scalar fields, though the proposed linear combination of the curvatures is interesting for tracking the effects of the projective symmetry. Now, since there is no symmetry that implies a unique nonminimal coupling function $f(\bm{\phi})$ for both curvature components (\ref{r_and_q}), an interesting dynamical aspects arise when the two curvature parts couple to matter via two distinct functions $f_{R}(\bm{\phi})$ and $f_{Q}(\bm{\phi})$, respectively. In this case, the invariant action that also satisfies the properties (i) to (iv) reads   
\begin{eqnarray}
\label{general_action2}
S[\Gamma, \bm{\phi}]=
\int d^{4}x \frac{\sqrt{ |f_{R}(\bm{\phi})R_{\mu\nu} + f_{Q}(\bm{\phi})Q_{\mu\nu}  - \partial_{\mu}\bm{\phi}^{a} \partial_{\nu}\bm{\phi}^{a}|}}
{V(\bm{\phi})} \nonumber \\
\end{eqnarray}

A remarkable feature of this action compared to the previous one is that the nonminimal couplings can be absorbed only from one of the curvatures, not both of them simultaneously (see Refs \cite{affine_inflation, entropy_production_in_affine_inflation} for how to make transition to minimal couplings in the symmetric models). Here, all the field equations of this action will emerge like in the previous section. Similarly, for simplicity one introduces the tensor field
\begin{eqnarray}
\label{tensor_k2}
K_{\mu\nu}(\Gamma,\bm{\phi})=
f_{R}(\bm{\phi})R_{\mu\nu} + f_{Q}(\bm{\phi}) Q_{\mu\nu}   - \partial_{\mu}\bm{\phi}^{a} \partial_{\nu}\bm{\phi}^{a}
\nonumber \\
\end{eqnarray}
The  equation of motion  in this case, arising from variation with respect to the connection, reads
\begin{widetext}
\begin{eqnarray}
\label{general_dynamical_equation2}
&&\bm{\nabla}_{\alpha} \left(f_{R}(\bm{\phi})\frac{\sqrt{|K|}}{V(\phi)} (K^{-1})^{\mu\nu} \right) 
- \bm{\nabla}_{\beta} \left(f_{R}(\bm{\phi})\frac{\sqrt{|K|}}{V(\phi)} (K^{-1})^{\mu\beta} \right)\delta^{\nu}_{\,\,\alpha} 
+2f_{R}(\bm{\phi})\frac{\sqrt{|K|}}{V(\bm{\phi})} (K^{-1})^{\mu\beta}\bm{S}_{\,\,\alpha\beta}^{\nu}
\nonumber \\
&&-2f_{R}(\bm{\phi})\frac{\sqrt{|K|}}{V(\bm{\phi})} (K^{-1})^{\mu\nu}\bm{S}_{\alpha}
+2f_{R}(\bm{\phi})\frac{\sqrt{|K|}}{V(\bm{\phi})} (K^{-1})^{\mu\beta}\bm{S}_{\beta} \delta^{\nu}_{\,\,\alpha} -2\partial_{\beta} \left(f_{Q}(\bm{\phi})\frac{\sqrt{|K|}}{V(\bm{\phi})} (K^{-1})^{[\nu\beta]} \right)\delta^{\mu}_{\,\,\alpha} =0.
\end{eqnarray}
\end{widetext}
which describes the purely affine version of the gravitational field equations involving high order terms in the connection which make them difficult to solve directly. Therefore, as in (\ref{metric_density}) the generalized metrical tensor density can be defined as
\begin{eqnarray}
\label{metric_density2}
M^{2}\sqrt{|\bm{g}|}\,\bm{g}^{\mu\nu} =
f_{R}(\bm{\phi})\frac{\sqrt{|K(\bm{\Gamma},\bm{\phi})|}}{V(\bm{\phi})} (K^{-1})^{\mu\nu},
\end{eqnarray}
Notice  that this density is proportional to $f_{R}(\bm{\phi})$ not $f_{Q}(\bm{\phi})$. The  reason is that in the Eddington gravity approach, the metric density emerges from the derivative of the Lagrangian with respect to the Ricci tensor $R_{\mu\nu}(\Gamma)$ as in (\ref{metric_as_derivative_wrt_ricci}), and since the latter can always have a symmetric part,  the physical symmetric metric is always defined from this approach. A tensor density defined from the derivative of the Lagrangian with respect to $Q_{\mu\nu}(\Gamma)$ cannot be symmetric, and hence cannot be used to define a physical metric.  

As we have seen in the previous section, in purely affine gravity the tensor density (\ref{metric_density}) or (\ref{metric_density2}) which is generated \textit{a posteriori} does not only lead to the concept of metric but it turns out to be very essential in simplifying the dynamics of the connection which arise in very complicated equations of motion.  
Given the metric tensor density (\ref{metric_density2}), the last equation of motion takes the form 
\begin{widetext}
\begin{eqnarray}
\label{metric_dynamical_equation2}
&&
\bm{\nabla}_{\alpha}(\sqrt{|\bm{g}|}\,\bm{g}^{\mu\nu})
+2\sqrt{|\bm{g}|}\,\bm{g}^{\mu\beta}\left(\bm{S}^{\nu}_{\,\,\alpha\beta} + \frac{1}{3}\bm{S}_{\beta} \delta^{\nu}_{\,\,\alpha} \right) 
-2\sqrt{|\bm{g}|}\,\bm{g}^{\mu\nu}\bm{S}_{\alpha} 
+2\partial_{\beta}\left(\frac{f_{Q}}{f_{R}}\right)
\left(\frac{1}{3}\sqrt{|\bm{g}|}\,\bm{g}^{[\mu\beta]}\,\delta^{\nu}_{\,\,\alpha}
-
\sqrt{|\bm{g}|}\,\bm{g}^{[\nu\beta]}\,\delta^{\mu}_{\,\,\alpha}\right) 
\nonumber \\
&&
+2\frac{f_{Q}}{f_{R}}\left(\frac{1}{3}\mathcal{J}^{\mu} \delta^{\nu}_{\,\,\alpha}- \mathcal{J}^{\nu} \delta^{\mu}_{\,\,\alpha}
\right) =0.
\end{eqnarray}
\end{widetext}
This implies that instead of (\ref{divergence_of_g}), one gets
\begin{eqnarray}
\label{divergence_of_g2}
\left(1 + \frac{4f_{Q}}{f_{R}}\right)\mathcal{J}^{\mu}
+4\partial_{\alpha}\left(\frac{f_{Q}}{f_{R}}\right)\sqrt{|\bm{g}|}\bm{g}^{[\alpha \mu]}
=0 
\end{eqnarray}
which means that unlike the previous model, the quantity $\mathcal{J}^{\mu}$ does not vanish, and therefore it influences the solution to the dynamical equation (\ref{metric_dynamical_equation2}). This is one of the effects of distinct couplings between the scalar fields and the Ricci tensors. Now, the identity (\ref{metric_density2}) represents the gravitational field equations of model (\ref{general_action2}), and it is equivalent to
\begin{eqnarray}
\label{field_equations2}
f_{R}(\bm{\phi})R_{\mu\nu} + f_{Q}(\bm{\phi})Q_{\mu\nu}  - \partial_{\mu}\bm{\phi}^{a} \partial_{\nu}\bm{\phi}^{a} 
-\frac{M^{2}V(\bm{\phi})}{f_{R}(\bm{\phi})} \bm{g}_{\mu\nu} =0 \nonumber \\
\end{eqnarray}

Another interesting property of the purely affine gravity with scalar fields is that the metric tensor can be integrated out easily from the field equations. This is clear from both equations (\ref{field_equations2}) and  (\ref{field_equations_1}) thanks to the nonzero potential. As we have seen above, $V(\bm{\phi})\neq 0$ is a primary requirement for generating the metric in this approach, and in the absence of the scalar fields (the case of free space), this metric will require a nonzero cosmological constant (already supported by observation) which replaces the potential \cite{inducing_gravity}. The idea of decoupling the metric tensor from matter fields through a metric-affine action in which the metric is not dynamical, has been used to construct a metric-free action for dark matter separable from ordinary matter sector \cite{affine_dark_matter}.   

As we have done in order to get the field equations (\ref{einstein_field_equations_1}), we also construct here a generalized Einstein tensor and write the previous equation in a standard form 
\begin{eqnarray}
\label{einstein_equations2}
&&
f_{R}(\bm{\phi})\left(R_{\mu\nu} - \frac{1}{2}\bm{g}_{\mu\nu}\bm{g}^{\alpha\beta} R_{\alpha\beta}\right) \nonumber \\
&&=
\partial_{\mu}\bm{\phi}^{a} \partial_{\nu}\bm{\phi}^{a}
-\frac{1}{2}\bm{g}_{\mu\nu}
\bm{g}^{\alpha\beta}
\partial_{\alpha}\bm{\phi}^{a} \partial_{\beta}\bm{\phi}^{a}
-\frac{M^{2}V(\bm{\phi})}{f_{R}(\bm{\phi})} \bm{g}_{\mu\nu} \nonumber \\
&&-f_{Q}(\bm{\phi})\left(Q_{\mu\nu}
-\frac{1}{2}\bm{g}_{\mu\nu}\bm{g}^{\alpha\beta} Q_{\alpha\beta}\right).
\end{eqnarray}

The final equation one derives from action (\ref{general_action2}) is the equation for the fields $\bm{\phi}^{a}(x)$. Variation with respect to $\bm{\phi}^{a}$ yields
\begin{eqnarray}
\partial_{\nu}&& \left(\frac{\sqrt{|K|}}{V(\bm{\phi})} K^{(\mu\nu)}\partial_{\mu} \bm{\phi}^{a} \right)
-\frac{\sqrt{|K|}}{V^{2}(\bm{\phi})}\frac{\partial V}{\partial \bm{\phi}^{a}} \nonumber \\
+&&\frac{1}{2}\frac{\sqrt{|K|}}{V(\bm{\phi})} K^{\mu\nu}
\left(\frac{\partial f_{R}}{\partial \bm{\phi}^{a}}R_{\mu\nu}
+\frac{\partial f_{Q}}{\partial \bm{\phi}^{a}} Q_{\mu\nu}\right) =0,
\end{eqnarray}

Like the gravitational field equations (\ref{general_dynamical_equation2}), this is also complicated and written in terms of the connection only. It is this equation that must describe the evolution of the scalar fields in a background endowed with a connection and not a metric. Once again, using the metric tensor density $\bm{g}$ of (\ref{metric_density2}) we get
\begin{eqnarray}
\label{box_phi2}
&&\frac{\partial_{\nu}\left(\sqrt{|\bm{g}|}\,\bm{g}^{(\mu\nu)} \partial_{\mu}\bm{\phi}^{a} \right)}{\sqrt{|\bm{g}|}}
-\frac{\partial V}{\partial \bm{\phi}^{a}} 
+\frac{1}{2}\frac{\partial f_{R}}{\partial \bm{\phi}^{a}}\,\bm{g}^{\mu\nu}R_{\mu\nu} \\
\nonumber
&&+\frac{1}{2}\frac{\partial f_{Q}}{\partial \bm{\phi}^{a}}\, \bm{g}^{\mu\nu}Q_{\mu\nu}
+ \bm{\Psi}(\bm{\phi}) =0,
\end{eqnarray}
where the last term is given by
\begin{eqnarray}
\bm{\Psi}(\bm{\phi})=
\left(1-\frac{M^{2}}{f_{R}} \right)\frac{\partial V}{\partial \bm{\phi}^{a}}
-\frac{1}{f_{R}}\frac{\partial f_{R}}{\partial \bm{\phi}^{a}}\bm{g}^{(\mu\nu)}
\partial_{\mu}\bm{\phi}^{a} \partial_{\nu}\bm{\phi}^{a} \nonumber \\
\end{eqnarray}

All the remarks and comments addressed above concerning  equations (\ref{einstein_field_equations_1}) and (\ref{box_phi_1}) can be brought here. The main difference however relies on the constraint (\ref{divergence_of_g2}) for the vector density $\mathcal{J}^{\mu}$ which does not vanish in this case. This clearly affects the dynamics of the affine connection which cannot take the form (\ref{solution_for_the_connection}). Therefore, we conclude that the interaction between the scalar fields and the second Ricci tensor does not only bring new effects at the level of the equations of motion, but also plays an important role in the solution of the dynamical equation for the connection. 

The dynamical aspects of the gravitational setup that we have discusseded throughout the above sections can be summarized as follows: (\textit{a}) The proposed action principles stand on the affine connection as the central geometric element (\textit{b}) They extend Eddington gravity by allowing the effects of both torsion and the skew-symmetric Ricci curvature (\textit{c}) Matter is incorporated as multiple scalar fields and are permitted to interact directly with the two Ricci tensors (\textit{d}) As a limiting case, the resulting dynamics is equivalent to that of GR when both connection and Ricci tensor are taken symmetric. When one or both of this symmetric characters are relaxed, the dynamics then shows significant difference from GR. Since it involves the scalar fields in the first place, this setup must be explored in line with the well-known scenarios, mainly inflation and dynamical dark energy. 

\section{Conclusion}
\label{section:conclusion}

In this paper we have explored various dynamical aspects of the asymmetic affine theory with scalar fields in the spirit of Eddington gravity. We have started with the fact that in the latter, the fundamental quantities behind the action principle, namely the connection and the associated curvature, are constrained to be symmetric in the first place leaving no place for other interesting geometric objects that can lead to new physical and cosmological effects. Furthermore, due to the absence of the concept of metric the action of the theory is known to be very arduous to accept matter fields. For those reasons we intended, firstly, to enlarge the theory by alleviating the constraints applied on the geometric quantities, and as a result, the theory gained new objects namely the torsion and the skew-symmetric part of the Ricci tensor. Secondly, we incorporated matter sources into the action as multiple scalar fields which seem not to require a metric. With this structure, we have investigated various possible ways in which the scalars are coupled to the central object (the affine connection) via the two types Ricci tensors, and have realized that crucial effects and deviations from general relativity can emerge from the nonminimal couplings between matter sources and both curvatures. 

The second Ricci tensor cannot be involved in the purely metric theories of gravity since it vanishes once the connection is metric compatible (Levi-Civita) from the beginning. Otherwise, this skew-symmetric part can enter the gravitational action but in the Palatini formulation where also an independent connection is introduced, and in this case the resulting theories may resemble that of a vector field \cite{ricci_as_vector_field}. In contrast, the present setup does not stand on the Palatini formulation. Indeed, the second Ricci curvature, one of the essential elements in this framework, need not to couple to metric since it is already a second rank tensor field that contributes to the volume measure itself.   

Throughout the paper, we have focused on the resulting equations of motion which originally take a purely affine form, and  provided a way to express them in a familiar form. It is through this step that the concept of metric emerges and facilitates the solution of the dynamical equations for the connection. At this point, we have shown that the solution (the connection in terms of the metric) can be similar to the vacuum case if the scalar fields are coupled to both curvatures by the same function, whilst different coupling functions lead to different physics. Another important feature in this framework is that due to the relaxation of the constraint on the curvature, the metric tensor of the theories emerges with both symmetric and anti-symmetric parts where the former is taken as the relevant physical field for the gravitational phenomenon whilst the latter contributes to the measurable deviations from the symmetric models.

Finally, one needs more investigations throughout the present aspects especially when these scalar fields are considered as sources of some cosmological phenomena. We leave this quest for separate study \cite{in_progress}.    

\section*{acknowledgements}
The work is supported by the United Arab Emirates University under UPAR Grant No. 12S004. The authors are thankful to Durmu\c{s} Demir and Gonzalo Olmo for discussing Eddington gravity.


\begin{thebibliography}{99}

\bibitem{unified_field_theory} 
  H.~F.~M.~Goenner,
  \textit{On the history of unified field theories},
  Living Rev.\ Rel.\  {\bf 7}, 2 (2004);
  H.~F.~M.~Goenner,
  \textit{On the History of Unified Field Theories. Part II. (ca. 1930 - ca. 1965)},
  Living Rev.\ Rel.\  {\bf 17}, 5 (2014).

\bibitem{eddington}
A. S. Eddington, \textit{The Mathematical Theory of Relativity}, Cambridge University Press (1923);
E. Schr\"{o}dinger, \textit{Space-Time Structure}, Cambridge University Press (1950).

\bibitem{einstein_nature}
A. Einstein, \textit{The Theory of the Affine Field}, Nature \textbf{112}, 448–449 (1923).

\bibitem{hlavaty}
V. Hlavaty, \textit{Geometry of Einstein's Unified Field Theory}, P. Noordhoff Ltd., Groningen, The Netherlands (1957).

\bibitem{lavinia_review}
F.~W.~Hehl, J.~D.~McCrea, E.~W.~Mielke and Y.~Ne'eman,
\textit{Metric affine gauge theory of gravity: Field equations, Noether identities, world spinors, and breaking of dilation invariance},
Phys. Rept. \textbf{258}, 1-171 (1995);
L.~Heisenberg,
\textit{A systematic approach to generalisations of General Relativity and their cosmological implications},
Phys. Rept. \textbf{796}, 1-113 (2019);
J.~Beltr\'an Jim\'enez, L.~Heisenberg and T.~Koivisto,
\textit{Coincident General Relativity},
Phys. Rev. D \textbf{98}, no.4, 044048 (2018);

\bibitem{cartan_kibble1}
T.~W.~B.~Kibble,
\textit{Lorentz invariance and the gravitational field},
J. Math. Phys. \textbf{2}, 212-221 (1961);
D.~W.~Sciama,
\textit{The Physical structure of general relativity},
Rev. Mod. Phys. \textbf{36}, 463-469 (1964);
F.~W.~Hehl, P.~Von Der Heyde, G.~D.~Kerlick and J.~M.~Nester,
\textit{General Relativity with Spin and Torsion: Foundations and Prospects},
Rev. Mod. Phys. \textbf{48}, 393-416 (1976).

\bibitem{cartan_kibble2}
A.~Trautman,
\textit{Einstein-Cartan theory},
[arXiv:gr-qc/0606062 [gr-qc]];
A.~Trautman,
\textit{Spin and Torsion May Avert Gravitational Singularities}, 
Nature Physical Science 242, 7–8 (1973);
N.~J.~Poplawski,
\textit{Nonsingular, big-bounce cosmology from spinor-torsion coupling},
Phys. Rev. D \textbf{85}, 107502 (2012);
S.~Desai and N.~J.~Pop\l{}awski,
\textit{Non-parametric reconstruction of an inflaton potential from Einstein\textendash{}Cartan\textendash{}Sciama\textendash{}Kibble gravity with particle production},
Phys. Lett. B \textbf{755}, 183-189 (2016).

\bibitem{bicak}
J. Bičák, \textit{On the Rainich geometrization of a vector meson field in the Kibble theory}, Czech J Phys 16, 95–98 (1966).

\bibitem{kijowski}
J.~Kijowski and R.~Werpachowski,
  \textit{Universality of affine formulation in general relativity theory},
  Rept.\ Math.\ Phys.\  {\bf 59}, 1 (2007);
  J. Kijowski, \textit{On a new variational principle in general relativity and the energy of the gravitational field}, Gen.Rel.Grav.,9,857 1978.
  
\bibitem{treder}
H.~H.~von Borzeszkowski and H.~J.~Treder,
 \textit{Spinorial matter in affine theory of gravity and the space problem},
  Gen.\ Rel.\ Grav.\  {\bf 33}, 1351 (2001).
\bibitem{demir_matter}
D.~A.~Demir,
\textit{Riemann-Eddington theory: Incorporating matter, degravitating the cosmological constant},
Phys. Rev. D \textbf{90}, no.6, 064017 (2014).

\bibitem{cota_matter}

J.~L.~Cervantes-Cota and D.~E.~Liebscher,
\textit{On constructing purely affine theories with matter},
Gen. Rel. Grav. \textbf{48}, no.8, 108 (2016);

\bibitem{azri_separate}
H.~Azri,
\textit{Separate Einstein-Eddington Spaces and the Cosmological Constant},
Annalen Phys. \textbf{528}, 404-411 (2016).

\bibitem{azri_immersed}
H.~Azri,
\textit{Eddington\textquoteright{}s gravity in immersed spacetime},
Class. Quant. Grav. \textbf{32}, no.6, 065009 (2015).

\bibitem{pad}
S.~Chakraborty and T.~Padmanabhan,
  \textit{Eddington gravity with matter: An emergent perspective},
  Phys.\ Rev.\ D {\bf 103}, no. 6, 064033 (2021).



\bibitem{affine_inflation}
H.~Azri and D.~Demir,
\textit{Affine Inflation},
Phys. Rev. D \textbf{95}, no.12, 124007 (2017);
H.~Azri and D.~Demir,
\textit{Induced Affine Inflation},
Phys. Rev. D \textbf{97}, no.4, 044025 (2018);
H.~Azri,
\textit{Are there really conformal frames? Uniqueness of affine inflation},
Int. J. Mod. Phys. D \textbf{27}, no.09, 1830006 (2018).


\bibitem{entropy_production_in_affine_inflation}
H.~Azri and S.~Nasri,
\textit{Entropy production in affine inflation},
Phys. Rev. D \textbf{101}, no.6, 064073 (2020).

\bibitem{affine_dark_matter}
H.~Azri, A.~Jueid, C.~Karahan and S.~Nasri,
  \textit{Affine gravitational scenario for dark matter decay},
  Phys.\ Rev.\ D {\bf 102}, no. 8, 084036 (2020).

\bibitem{scalar_connection_gravity} 
  H.~Azri and S.~Nasri,
  \textit{Scalar-Connection Gravity and Spontaneous Scalarization},
  Phys.\ Rev.\ D {\bf 103}, no. 2, 024035 (2021).


\bibitem{inducing_gravity} 
  H.~Azri,
  \textit{Inducing Gravity From Connections and Scalar Fields},''
  Class.\ Quant.\ Grav.\  {\bf 36}, no. 16, 165006 (2019)


\bibitem{lovelock}
D. Lovelock and H. Rund,
\textit{Tensors, Differential Forms and Variational Principles},  Dover Publications Inc (1990).

\bibitem{azri_thesis} 
  H.~Azri,
  \textit{Cosmological Implications of Affine Gravity},
  arXiv:1805.03936 [gr-qc].

\bibitem{murphy} 
  G.~L.~Murphy,
  \textit{Affine-Projective Field Laws},
  Phys.\ Rev.\ D {\bf 11}, 2752 (1975).
  
\bibitem{poplawski}
N.~J.~Poplawski,
\textit{On the nonsymmetric pure-affine gravity},
Mod. Phys. Lett. A \textbf{22}, 2701-2720 (2007).

\bibitem{multifield_inflation}
K.~A.~Malik and D.~Wands,
  \textit{Adiabatic and entropy perturbations with interacting fluids and fields},
  JCAP {\bf 0502}, 007 (2005);
  D.~Langlois and B.~van Tent,
  \textit{Hunting for Isocurvature Modes in the CMB non-Gaussianities},
  Class.\ Quant.\ Grav.\  {\bf 28}, 222001 (2011);
  D.~Langlois and B.~van Tent,
  \textit{Isocurvature modes in the CMB bispectrum},
  JCAP {\bf 1207}, 040 (2012);
  


\bibitem{kaiser} 
  D.~I.~Kaiser and A.~T.~Todhunter,
  \textit{Primordial Perturbations from Multifield Inflation with Nonminimal Couplings},
  Phys.\ Rev.\ D {\bf 81}, 124037 (2010);
  D.~I.~Kaiser,
  \textit{Nonminimal Couplings in the Early Universe: Multifield Models of Inflation and the Latest Observations},
  Fundam.\ Theor.\ Phys.\  {\bf 183}, 41 (2016);
  K.~A.~Malik and D.~Wands,
  \textit{Adiabatic and entropy perturbations with interacting fluids and fields},
  JCAP {\bf 0502}, 007 (2005);
  J.~White, M.~Minamitsuji and M.~Sasaki,
  \textit{Curvature perturbation in multi-field inflation with non-minimal coupling},
  JCAP {\bf 1207}, 039 (2012);
  P.~Carrilho, D.~Mulryne, J.~Ronayne and T.~Tenkanen,
  \textit{Attractor Behaviour in Multifield Inflation},
  JCAP {\bf 1806}, 032 (2018).

\bibitem{ricci_as_vector_field}
D.~Demir and B.~Puli\c{c}e,
\textit{Geometric Dark Matter},
JCAP \textbf{04}, 051 (2020);
A.~T.~Filippov,
\textit{Affine generalizations of gravity in the light of modern cosmology},
Proc. Steklov Inst. Math. \textbf{272}, no.1, 107-118 (2011);
N.~J.~Poplawski,
\textit{On the Maxwell Lagrangian in the purely affine gravity},
Int. J. Mod. Phys. A \textbf{23}, 567-579 (2008).

  
\bibitem{in_progress}
Work in progress (2021).

\end{thebibliography}
\end{document}